\documentclass[fleqn,usenatbib]{mnras}

\usepackage{newtxtext,newtxmath}

\usepackage[T1]{fontenc}
\usepackage{ulem}

\DeclareRobustCommand{\VAN}[3]{#2}
\let\VANthebibliography\thebibliography
\def\thebibliography{\DeclareRobustCommand{\VAN}[3]{##3}\VANthebibliography}

\usepackage{graphicx}	
\usepackage{amsmath}	
\usepackage{xcolor}

\usepackage[acronym]{glossaries}
\makeglossaries
\newacronym{GW}{GW}{gravitational wave}
\newacronym{NS}{NS}{neutron star}
\newacronym{BH}{BH}{black hole}
\newacronym{BBH}{BBH}{binary black hole}
\newacronym{BNS}{BNS}{binary neutron star}
\newacronym{BHNS}{BHNS}{black hole-neutron star}
\newacronym{EoS}{EoS}{equation of state}
\newacronym{EM}{EM}{electromagnetic}
\newacronym{sGRB}{sGRB}{short Gamma-Ray Burst}
\newacronym{ISCO}{ISCO}{Innermost Stable Circular Orbit}
\newacronym{CTT}{CTT}{Conformal Transverse-Traceless}
\newacronym{CTS}{CTS}{Conformal Thin Sandwich}
\newacronym{xCTS}{xCTS}{extended Conformal Thin-Sandwich}
\newacronym{ID}{ID}{initial data}
\newacronym{PDE}{PDE}{partial differential equation}
\newacronym{NR}{NR}{numerical relativity}
\newacronym{GR(M)HD}{GR(M)HD}{General Relativistic (Magneto) Hydrodynamics}
\newacronym{HRSC}{HRSC}{High-Resolution Shock-Capturing}
\newacronym{AMR}{AMR}{Adaptive Mesh Refinement}
\newacronym{GRHD}{GRHD}{general relativistic hydrodynamics}
\usepackage{siunitx}
\DeclareSIUnit{\erg}{erg}
\DeclareSIUnit{\Mass}{\mathit{M}}
\DeclareSIUnit{\c}{\mathit{c}}
\DeclareSIUnit{\dyn}{dyn}
\DeclareSIUnit{\Gauss}{G}
\DeclareSIQualifier{\Sun}{\ensuremath{\odot}}
\DeclareSIQualifier{\disk}{disk}
\DeclareSIQualifier{\BH}{BH}
\DeclareSIQualifier{\NS}{NS}
\DeclareSIUnit\clight{\text{\ensuremath{c}}}

\usepackage[capitalise,noabbrev]{cleveref}

\title{Impact of black hole spin on low-mass black hole--neutron star mergers}

\author[R. Matur, I. Hawke and N. Andersson]{
Rahime Matur\thanks{E-mail: r.matur@soton.ac.uk},
Ian Hawke,
Nils Andersson
\\
Mathematical Sciences and STAG Research Centre University of Southampton, Southampton SO17 1BJ, UK\\
}

\date{Accepted XXX. Received YYY; in original form ZZZ}

\pubyear{\the\year{}}

\begin{document}
\glsdisablehyper
\label{firstpage}
\pagerange{\pageref{firstpage}--\pageref{lastpage}}
\maketitle

\begin{abstract}

The recent detection of GW230529 suggests that black hole–neutron star mergers may involve low-mass black holes, potentially producing detectable electromagnetic counterparts. Motivated by this, we perform eleven fully general-relativistic hydrodynamic simulations with and without neutrino treatment, targeting the inferred chirp mass of GW230529. We systematically vary the black hole spin from $a_{\mathrm{BH}} = 0.0$ to $0.8$ in steps of $0.1$, making this the most comprehensive study of spin effects in black hole-neutron star mergers to date. We confirm our earlier findings of fast-moving ejecta ($v \geq 0.6\,c$) in this parameter regime and demonstrate a clear spin dependence, with fast-ejecta masses reaching up to $\qty{\sim 3e-4}{\Mass\Sun}$ for $a_{\mathrm{BH}} = 0.8$.
Most notably, we identify for the first time the presence of spiral wave–driven ejecta in black hole-neutron star mergers -- a phenomenon previously reported only in binary neutron star systems. The mass of this component grows significantly with spin, reaching levels up to $\qty{\sim 7e-3}{\Mass\Sun}$. These results establish a new spin-enhanced mechanism for powering blue kilonova emission in black hole-neutron star mergers, significantly extending the range of systems expected to produce observable electromagnetic counterparts.
\end{abstract}

\section{Introduction}

Multi-messenger events, involving observations of both gravitational waves and electromagnetic signals, offer a more complete picture of the underlying physics. A landmark example of this was the first confirmed multi-messenger event; the binary neutron star merger GW170817~\citep{2017PhRvL.119p1101A, grb} (with electromagnetic counterparts AT2017gfo and GRB170817A~\citep{doi:science_em170817, multimessenger1}). This event confirmed the promise of multi-messenger astronomy and provided a wealth of astrophysical insight. It placed constraints on the tidal deformability of neutron stars, setting an upper bound of $\Lambda \leq 800$. Furthermore, the associated short gamma-ray burst, GRB170817A~\citep{gammaray, integral}, was observed just $\qty{\sim 1.7}{\second}$ after the merger by the Fermi Gamma-ray Burst Monitor and INTEGRAL. This near-simultaneity offered a unique opportunity to test the propagation speed of gravitational waves, confirming consistency with the speed of light and providing a validation of general relativity~\citep{grb}. 

In contrast, for black hole-neutron star mergers the occurrence of tidal disruption---and thus the likelihood of an electromagnetic counterpart---depends on several key parameters (see~\cite{reviewfoucart, reviewkyutoku} for recent reviews and references to the relevant literature). These include a low mass ratio, defined as
\begin{equation}
    Q = \frac{M_{\mathrm{BH}}}{M_{\mathrm{NS}}},
\end{equation}
where $M_{\mathrm{BH}}$ and $M_{\mathrm{NS}}$ are the gravitational masses of the black hole and neutron star, respectively; a spinning black hole, characterised by the dimensionless spin parameter
\begin{equation}
    a_{\mathrm{BH}} = \frac{c\, J_{\mathrm{BH}}}{G\, M_{\mathrm{BH}}^2},
\end{equation}
where $J_{\mathrm{BH}}$ is the angular momentum of the black hole, \(c\) is the speed of light, and \(G\) is the gravitational constant; and a low neutron star compactness, given by
\begin{equation}
    \mathcal{C} = \frac{G\, M_{\mathrm{NS}}}{c\,^2 R_{\mathrm{NS}}},
\end{equation}
where $R_{\mathrm{NS}}$ is the radius of the neutron star.

Even though no multi-messenger observation has yet been confirmed for observed black hole-neutron star mergers~\citep{twobhns, 2024ApJ...970L..34A}, the recent detection of the GW230529 event~\citep{2024ApJ...970L..34A} has renewed interest in the tidal disruption scenario. This event is particularly interesting due to its low mass ratio, which makes it a promising candidate for multi-messenger observation of black hole-neutron star mergers, even though such prospects strongly depend on the neutron star equation of state and the black hole’s spin.

To investigate the complex physics underlying multi-messenger events, numerical relativity simulations are the only available tool. Such simulations allow one to model the emission of gravitational waves, neutrino transport, magnetic field effects, rapid neutron capture process ($r$-process) nucleosynthesis, and other complex physical phenomena. While state-of-the-art numerical relativity studies can capture many of these aspects, albeit in an approximate manner, there are limitations, including systematic errors due to  artificial heating during the inspiral~\citep{problematic}, the absence of realistic magnetic field configurations~\citep{2025arXiv250618995G}, approximations made in the treatment of neutrino transport~\citep{foucartneutrinoreview}, and uncertainties in microphysical modelling. 

To address these limitations, efforts tend to advance either the physical realism or the numerical methods used. For instance, \citet{nmesh} introduced the discontinuous Galerkin method for general relativistic hydrodynamics. To improve neutrino transport, \citet{foucart20_montecarlo, Foucart23_mc_bh} performed the first binary neutron star merger simulations with Monte-Carlo neutrino transport. A merger involving a subsolar-mass black hole was explored in \citet{dietrichsubsolar}. Long-term self-consistent simulations of black hole-neutron star mergers~\citep{hayashi2022tab} and binary neutron star mergers~\citep{Kiuchi_23_secondslongBNS}, extending up to $\qty{\sim 2}{\s}$ were presented. \citet{Kiuchi_24_mag} showed that magnetars formed in binary neutron star mergers can launch jets and power bright kilonovae. \citet{radiceneutrinooscillation} explored the role of neutrino flavour transformation in binary neutron star mergers and demonstrated its impact on the composition and nucleosynthesis of the ejecta. More recently, jet formation following accretion-induced prompt black hole formation in a binary neutron star merger has been explored up to $\qty{\sim 1.5}{\s}$ after the merger~\citep{Hayashi_25_jetprompt}. However, there is still much work to be done to ensure that these simulations become more realistic.

Black hole–neutron star mergers involving spinning black holes have been extensively explored in the literature. Many studies have considered systems with spin using simple equations of state~\citep{etienne2009, kyutoku2009, duez2009, foucart2010, kyutoku2011, foucart2011, east2011, etienne2011, foucart2013, lackey2014, paschalidis2015, kyutoku2015, kawaguchi2015, kuichi2015, wan2017, ruiz2018, foucart2019, hayashi2021, foucart2020, PhysRevD.104.084010, gottlieb2023, izquierdo2024, chen2024}, while more recent efforts have employed finite-temperature, composition-dependent matter models~\citep{foucart2014tab, foucart2015tab, deaton2016tab, foucart2016tab, kyutoku2016tab, brege2018tab, desai2019tab, foucartlowmass, most2020tab, hayashi2022tab, hayashi2023tab, most2023tab, martineau2024tab, matur2024tab, topolski2024tab}.

\cite{east2015} investigated spinning neutron stars---rather than spinning black holes---reporting that neutron star spin increases both bound and ejecta masses. Meanwhile,~\cite{eastdarkmatter, ruiz2021} explored configurations involving spins on both the neutron star and the black hole. However, even though there are plenty of work that models black holes with spin, only a small number of studies simulate more than a few different spin values~\citep{lackey2014, ruiz2018, most2023tab, martineau2024tab, topolski2024tab} with a typical spin increment of $0.2$ for spins aligned with the orbital angular momentum. These studies report that including the black hole spin results in more violent mergers, and the final black hole being surrounded by more dense matter. 

To provide a more general picture,~\cite{foucartremnant} analysed approximately $75$ black hole–neutron star merger simulations, covering both spinning and non-spinning configurations across a range of mass ratios, using three different numerical codes. More recently,~\cite{AlejandraSpin} presented $52$ simulations exploring the effects of equation of state, mass ratio, and spin orientation. These studies provide valuable insights into black hole-neutron star mergers, but they do not isolate the role of black hole spin. 
Furthermore, although some studies have investigated low-mass black hole–neutron star mergers~\citep{foucartlowmass, hayashi2021, dietrichsubsolar, matur2024tab, dietrich230529}, this regime  requires further exploration. In particular, a direct and controlled investigation of the impact of black hole spin in low mass-ratio systems remains outstanding.

While most simulations of spinning black hole–neutron star mergers in the literature focus on relatively high mass ratios and employ simplified equations of state---such as piecewise polytropic or gamma-law models---only a few studies have explored systems with lower mass ratios and a range of spin values~\citep{kyutoku2011, ruiz2018, chen2024}. 

\citet{kyutoku2011} performed simulations with three spin configurations ($a_{\mathrm{BH}} = -0.5$, $0.5$, and $0.75$) with a mass ratio of $Q = 3$. They found that an aligned black hole spin enhances tidal disruption through spin–orbit coupling, extends the disc to larger radii due to the reduced ISCO radius, alters the final spin of the remnant black hole, and increases the number of gravitational wave cycles before merger  (consistently with~\cite{etienne2009}). 

\citet{ruiz2018} performed four simulations targeting the same mass ratio and spins as in ~\citet{kyutoku2011}. Their results show that only in the case of high spin is a relativistic jet successfully launched, highlighting the importance of spin for jet formation. More recently, ~\citet{chen2024} performed simulations with black hole spins of $a_{\mathrm{BH}} = 0.5$ and $0.75$ for a mass ratio of $Q = 3$, reporting that both the baryonic remnant disc mass and the ejecta amount increase with higher black hole spin. They also noted that, for some configurations the average velocity of the ejecta was significantly enhanced with increasing spin. Even though all of these studies have significantly contributed to the field, a more systematic investigation, particularly for low mass ratio black hole–neutron star systems, remains a crucial missing piece.

To address this gap, we perform $11$ spinning black hole-neutron star merger simulations with a low mass ratio ($Q = 2.6$), including $9$ models with neutrino transport and $2$ without. In the first $9$ simulations, we vary the dimensionless spin parameter of the black hole from $0$ to $0.8$ in increments of $0.1$, while keeping all other parameters fixed, such as the individual masses and the neutron star equation of state. This represents the most systematic investigation of spin effects in black hole–neutron star mergers, particularly in scenarios involving tidal disruption, to date. A recent study by~\cite{qinbhns} shows that, for systems with masses similar to GW230529, mass accretion during binary evolution can spin black holes up to about $0.65$. This supports our choice of spins in this analysis. In this study, while mostly focusing on the post-merger phase, we examine how the baryonic remnant mass, various ejecta components (fast-moving ejecta, spiral wind-driven ejecta etc.), and the evolution of density modes are influenced by the black hole spin. We also show how the abundances of heavy elements change due to nucleosynthesis.

The paper is organized as follows. In~\cref{sec:numericalmethods}, we describe the initial configurations and the numerical methods used in our simulations. In~\cref{sec:results}, we present the main results, including the properties of the different ejecta components---particularly the spiral wind-driven ejecta and $r$-process nucleosynthesis. Finally, in~\cref{sec:conclusion}, we summarize our findings and conclude the paper.

\section{Numerical Setup}\label{sec:numericalmethods}

In general relativistic hydrodynamics simulations of \gls{BNS} and \gls{BHNS} mergers, spacetime and hydrodynamics parts are treated separately. The central approach involves discretising space into cells and evolving the relevant equations forward in time based on \(3{+}1\) decomposition. For the spacetime evolution, we solve the Einstein Field Equations (EFE), which are a set of nonlinear partial differential equations. Within the \gls{NR} framework, these equations are decomposed into constraint and evolution parts. The constraint equations are used in constructing the \gls{ID} by applying methods such as the \gls{xCTS}~\citep{1999PhRvL..82.1350Y, 2003PhRvD..67d4022P} method, while the evolution equations govern the dynamical evolution of the spacetime.

We use the \texttt{FUKA} \gls{ID} solver~\citep{fuka}, which uses the \gls{xCTS} method, to construct the \gls{ID}. The mass ratio is fixed to $Q = 2.6$, with gravitational masses of the \gls{BH} and \gls{NS} set to $\qty{3.6}{M_\odot}$ and $\qty{1.4}{M_\odot}$, respectively. This yields a chirp mass of $\qty{1.91}{M_\odot}$, computed as $(M_{\mathrm{BH}} M_{\mathrm{NS}})^{3/5}/(M_{\mathrm{BH}} + M_{\mathrm{NS}})^{1/5}$, which matches that of GW230529. We perform $9$ simulations including neutrino transport, systematically varying the dimensionless spin parameter of the \gls{BH} from $a_{\mathrm{BH}} = 0$ to $0.8$ in increments of $0.1$, with all spins aligned with the orbital angular momentum. The models are labelled as \texttt{Q2.6aX}, where $X$ denotes the black hole spin parameter $a_{\mathrm{BH}}$ in the range $0 \leq a_{\mathrm{BH}} \leq 0.8$. Additionally, we perform $2$ simulations without neutrino transport for the \texttt{Q2.6a0} and \texttt{Q2.6a08} models. 

While creating the \gls{ID}, \texttt{FUKA} divides the computational domain into three spherical regions around each star which represent a nucleus and two spherical-like shells~\citep{fuka}. In our case, we use $(N_r, N_\theta, N_\phi) = (9, 9, 8)$ collocation points per domain in the radial, polar, and azimuthal directions, respectively.

For all models, we employ the finite-temperature, composition-dependent \texttt{DD2} \gls{EoS}~\citep{2012ApJ...748...70H} available from \href{https://stellarcollapse.org}{StellarCollapse}~\citep{stellarcollapse}. Our goal is to explore how the physical observables---such as the ejecta properties, remnant disc mass, and neutrino energies and luminosities---depend on the black hole spin in low-mass-ratio \gls{BHNS} mergers within a mass range relevant to a real observation.

We use a cell-centred grid structure that extends to \qtylist[list-units=bracket,list-final-separator={, }]{2835;2835;1418}{\km} with reflection symmetry on the $z$-axis. The \texttt{Carpet} \gls{AMR} driver~\citep{carpet} of \texttt{Cactus}~\citep{cactus} is used, with $7$ refinement levels, the finest of which has a resolution of \qty{221}{\m}. The finest refinement levels are centred around the compact objects: one around the neutron star, covering a radius of approximately $\qty{30}{km}$, and another around the black hole, covering a radius of \qty{15}{\km}, which is well beyond the apparent horizon radius of \qty{\sim 4.5}{km} for the non-spinning model. The extended refinement region around the black hole is intended to suppress unphysical recoil velocities. We observe that increasing the innermost refinement region contributes to a more stable spacetime evolution throughout the simulation.

Along with the main simulations perform at a resolution of \qty{221}{\m}, we conduct three additional runs to perform a convergence test. We select the \texttt{Q2.6a0} and \texttt{Q2.6a08} models for this purpose. Hereafter, we refer to LR, MR, and HR as the low-, medium-, and high-resolution simulations, respectively, corresponding to resolutions of \qty{346}{\m}, \qty{276}{\m}, and \qty{221}{\m}. In the main text, we discuss only the main (HR) simulation results, and therefore omit the “HR” label for clarity. Accordingly, we performed LR, MR, and HR runs for the \texttt{Q2.6a0} model, and MR and HR runs for the \texttt{Q2.6a08} model.

The spacetime evolution is carried out using \texttt{CTGamma}~\citep{ctgamma} that employs the \texttt{Z4c} formulation~\citep{z4c} of the EFE to have better control constraint violations. We also use several components of the \texttt{Sophie Kowalevski} release~\citep{2022zndo...7245853H} of the~\href{https://einsteintoolkit.org/} {\texttt{Einstein Toolkit}}~\citep{etcode}.

The spacetime evolution is performed using fourth-order finite-difference methods for spatial derivatives and a fourth-order Runge-Kutta (RK4) method for time integration with a \texttt{Courant-Friedrichs-Lewy} (CFL) factor of $0.15$. We use the \texttt{1+log} slicing condition for the lapse function and the \texttt{Gamma-driver} condition for the shift vector. The constraint damping parameters are set to $\kappa_{1} = 0.02$ and $\kappa_{2} = 0.0$.

For the hydrodynamical part, we have a set of conservation equations, particularly the conservation of the baryon number density and the energy-momentum tensor for the pure \gls{GRHD}. In the case of neutrino transport, the conservation of the energy-momentum tensor changes slightly~\citep{radice_2016_m0leakage},
\begin{equation}
    \nabla_{\beta} T^{\alpha \beta} = \Psi^{\alpha} = Q u^{\alpha}
\end{equation}
where $\Psi^{\alpha}$, $Q$ and $u^{\alpha}$ are the source term for the weak interactions, net neutrino cooling/heating rate per-unit volume and (instantaneous) four-velocity, respectively.

We use \texttt{WhiskyTHC}~\citep{2012A&A...547A..26R, 2014CQGra..31g5012R, 2014MNRAS.437L..46R, 2015ASPC..498..121R} for the hydrodynamical evolution. \texttt{WhiskyTHC} implements the evolution equations in flux-conservative form using the Valencia formulation~\citep{1991PhRvD..43.3794M, 1997ApJ...476..221B, 1999astro.ph.11034I} and solves them with a finite-volume method. This method divides the computational domain into discrete cells and evolves the cell-averaged values of the hydrodynamical variables. Discontinuities at cell interfaces---arising from differences between left and right states---are handled using the \texttt{Local Lax-Friedrichs} (LLF)~\citep{llf} flux-splitting method. For the reconstruction, we use the $5^{ \mathrm{th}}$-order monotonicity-preserving \texttt{MP5}~\citep{SURESH199783} method within \texttt{WhiskyTHC}.

For the neutrino transport, it would be ideal to solve a six-dimensional Boltzmann equation~\citep{foucartneutrinoreview}. However, this is computationally prohibitive in merger simulations. Most hydrodynamical codes therefore employ approximate transport schemes. In this study, we use the \texttt{M0+Leakage} scheme implemented in \texttt{WhiskyTHC}~\citep{radice_2016_m0leakage}, which can approximately capture the neutrino absorption in the baryon remnant~\citep{foucartneutrinoreview}. This scheme calculates the effective emissivity, number density, and average energy of free-streaming neutrinos for electron neutrinos ($\nu_e$), electron antineutrinos ($\bar{\nu}_e$), and heavy-lepton neutrinos ($\nu_x$)~\citep{radice_2016_m0leakage}. We use the same grid setup and follow the same procedure as in our previous work~\citep{2026MNRAS.545f2009K} to compute the neutrino energy and luminosities from the outer boundary, at $\qty{\sim 756}{\km}$, of the uniform spherical grid. We refer the reader to \citet{radice_2016_m0leakage} for further details of the neutrino transport scheme.

The ejecta properties are calculated using \texttt{Outflow} from a surface located at $\qty{\sim 300}{\km}$. As discussed by~\citet{kastaunejectedmass, 2021PhRvD.104l3010F}, the two most commonly used criteria to identify whether a fluid element is gravitationally unbound in neutron star merger simulations are the geodesic and Bernoulli conditions. In the geodesic criterion, a fluid element is considered unbound if $u_{t} < -1$, whereas the Bernoulli criterion requires $h u_{t} < -1$, where $u_{t}$ is the time component of the four-velocity and $h$ is the specific enthalpy. We calculate the dynamical ejecta according to the geodesic criterion. A part of the dynamical ejecta that moves with velocities of $v \geq 0.6\,c$ is classified as fast-moving~\citep{massejection}, while matter that does not have enough energy to become unbound remains bound to the system and is referred to as fallback matter. To quantify the spiral wave-driven component, we follow the approach of~\cite{spiral1}, computing the additional ejecta mass based on the Bernoulli criterion, starting from the moment when the dynamical ejecta saturates.

In addition to its mass, the composition of the ejecta, particularly the electron (or proton) fraction, $Y_{\mathrm{e}}$, defined as
\begin{equation}
    Y_{\mathrm{e}} = \frac{n_p}{n_p + n_n},
\end{equation}
where $n_p$ and $n_n$ are the number densities of protons and neutrons, respectively, plays a key role in the production of heavy elements via $r$-process nucleosynthesis.

The disc mass is calculated from three-dimensional snapshots by interpolating \gls{AMR} data onto a uniform grid. We calculate
\begin{equation}\label{eq:disc}
    M_{\mathrm{disc}} = \int \rho W \sqrt{\gamma}\, \mathrm{d}^3x,
\end{equation}
where $\rho$, $W$, and $\gamma$ are the rest-mass density of the fluid, the Lorentz factor, and the determinant of the spatial metric, respectively\footnote{We use the \href{https://bitbucket.org/dradice/scidata/}{Scidata} library to compute the disc masses.}.

Here, we emphasise that fallback material and the disc are distinct concepts. The fallback material is computed in the same way as the ejecta mass: we place a detector at a given distance and identify bound matter based on its energy. However, the disc mass is calculated as described in~\cref{eq:disc}.

We compute the evolution of the density modes following the standard procedure described in~\cite{spiral2}. Specifically, we consider six modes, with mode numbers $m=1$ to $6$. The corresponding mode amplitudes are calculated as~\citep{spiral2}
\begin{equation}
    C_{m}=\int_{z=0} e^{- \mathrm{i} m \varphi}\, \rho W \sqrt{\gamma}\, \mathrm{d}^2x
\end{equation}
We normalize the mode amplitudes by $C_0$, which corresponds to the $m = 0$ mode at the time of merger which in turn corresponds to the time when the gravitational wave amplitude reaches its maximum. While calculating the mode amplitudes, we take the recoil of the \gls{BH} into account.

To calculate the $r$-process nucleosynthesis, we need tracer particles that are initially located within the \gls{NS}. However, because there is a lack of tracer particles at the start of our simulations, we compute the tracer particle trajectories and the required hydrodynamical properties for the nucleosynthesis by performing a post-processing procedure. For the crucial first step of obtaining the trajectories, we use the three-dimensional snapshots of the three-velocity of selected fluid elements (which is measured by the normal observer) and integrate backwards in time to calculate the particle trajectories, following a similar approach to that used in supernova simulations~\citep{reichert1, sieverding}. The key point here is that these fluid elements should be unbound so they can undergo $r$-process nucleosynthesis. Therefore, while calculating the trajectories, we apply the geodesic criterion, as we do in the ejected mass calculation. Using these trajectories, we determine the rest-mass density, temperature, and electron fraction of the tracer particles. Finally, we use this data to calculate the nucleosynthesis with \texttt{WinNet}.

In summary, after the trajectories are obtained, the post-processing begins by using some of the input parameters to solve a set of ordinary differential equations (ODEs). Depending on the temperature regime, either the ``Network'' or the ``nuclear statistical equilibrium'' (NSE) modules are selected. For our purposes, we use the NSE module. In the NSE approach, for a given set of hydrodynamical quantities, \texttt{WinNet} solves three equations: (i) the Saha equation, which is derived by introducing the relevant chemical potentials of the nucleus, (ii) the mass conservation equation, and (iii) the charge neutrality equation. For further details on \texttt{WinNet}, we refer the reader to \cite{winnet}. To model the $r$-process nucleosynthesis from our simulations, we adapted the parameter file originally used in~\cite{RosswogEM1, RosswogEM2, RosswogEM3}.

\section{Results}\label{sec:results}

In discussing the results of our simulations, we focus on the post-merger phase by analyzing the properties of the ejected matter and the impact of neutrinos.

\subsection{System dynamics and ejected matter properties} 

\begin{figure*}
\centering
\includegraphics[width=\textwidth]{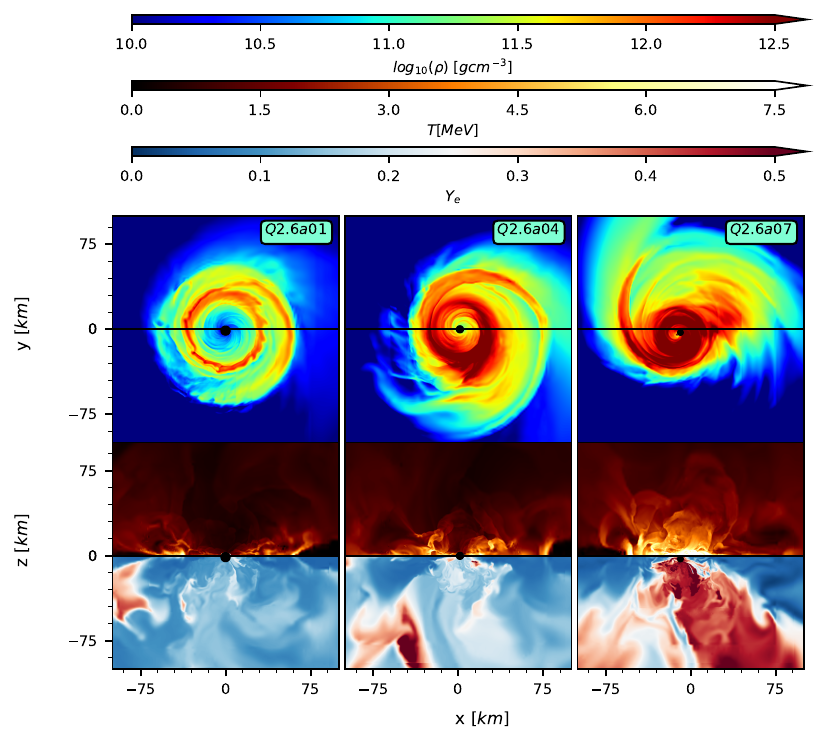} 
\caption{\label{fig:2d}Snapshots of the rest-mass density, temperature, and electron fraction $\qty{7}{\ms}$ after the merger for three models: \texttt{Q2.6a0}, \texttt{Q2.6a04} and \texttt{Q2.6a08}, respectively. The rest-mass density distributions indicate that all models undergo tidal disruption, with the highest-spin model exhibiting the most violent merger, leaving behind more massive and denser matter surrounding the black hole. This interpretation is further supported by the temperature snapshots, which show that the baryonic matter is significantly hotter in the highest-spin model compared to the other cases. At this stage, the electron fraction also tends to increase with increasing spin.}
\end{figure*}

We define the \gls{NS} as tidally disrupted if, in the post-merger phase, the maximum rest-mass density within a coordinate sphere of radius $\qty{150}{\kilo\metre}$ centered on the \gls{BH} remnant exceeds $\qty{6.2e6}{\gram\per\centi\metre\cubed}$. This corresponds to three orders of magnitude above the atmosphere density. \cref{fig:2d} shows that the maximum rest-mass density of this matter is significantly higher than this threshold, reaching at least \qty{e11}{\gram\per\centi\metre\cubed}. According to this definition, tidal disruption occurs in all our simulations, which is expected for systems with such small mass ratios~\citep{foucartlowmass, matur2024tab,martineau2024tab}. As anticipated, the severity of the disruption increases with higher black hole spin, resulting in denser and hotter baryonic matter remaining outside the black hole, as shown in~\cref{fig:2d}.~\footnote{The rest-mass density plots are produced using \texttt{PyCactus}~\citep{pycactus}.} To better quantify this effect, we calculate the properties of the ejected matter and the disc. As described in~\cref{sec:numericalmethods}, we extract ejecta properties at radial distances of $\qty{150}{\km}$ and $\qty{300}{\km}$. Since the matter does not reach the detector immediately after being ejected, tracking its radial velocity is crucial. For each detector, we account for the corresponding retarded time and then evaluate the ejecta properties accordingly.

A summary of the properties of the different ejecta components is presented in~\cref{table:finalproperties300}. In the following discussion, we concentrate on the properties measured at a radius of $\qty{300}{\km}$.

Before proceeding, we would like to mention the convergence test results for the  \texttt{Q2.6a0} and \texttt{Q2.6a08} models. While the error for \texttt{Q2.6a0} is consistent with previous work~\citep{massejection}, the error remains large for the \texttt{Q2.6a08} model. In particular, the errors in the total dynamical ejecta mass and the fast-moving ejecta are significant, whereas $Y_{e}$ and the mass of the fallback material show convergence. Therefore, we exclude this outlier model, \texttt{Q2.6a08}, from the relevant plots.

Looking at the mass-weighted $Y_{\mathrm{e}}$, we find that the relative error in the \texttt{Q2.6a0} model is at most $\sim 4.6$ per cent, whereas the relative error in the \texttt{Q2.6a08} model is negligible. Therefore, the most reliable parameter in our simulations is the mass-weighted $Y_{\mathrm{e}}$. Similarly, the mass-weighted velocity also converges with resolution, showing only $\sim 8$ per cent relative error. The relative error in the mass of the fallback material is at most $25$ per cent in both models.

The relative error of the dynamical ejecta for the \texttt{Q2.6a0} model is approximately 45 per cent, which is comparable to the relative error reported in~\cite{massejection}. For the fast-moving ejecta, the relative error increases to 65 per cent. These values are reasonable given the very small ejecta masses. More robust results will require higher resolution simulations, which we leave for future work.

\begin{figure*}
\centering
\includegraphics[width=\textwidth]{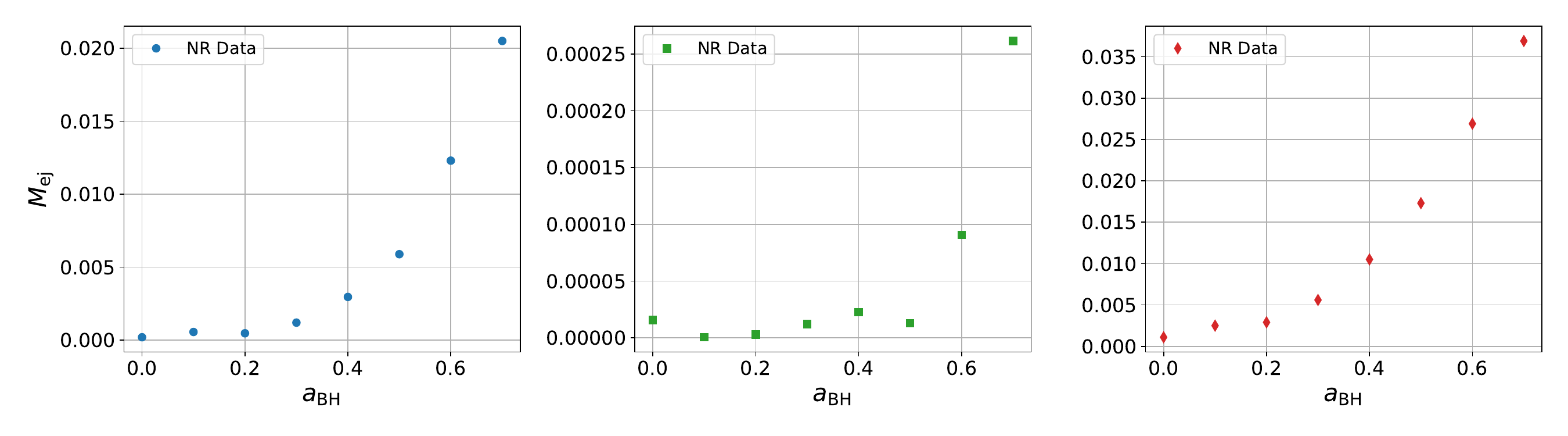}
\caption{\label{fig:compsejecta200}The relationship between the mass of the different ejecta components and spin. The left, central, and right panels show the variation of the total ejecta mass, fast-moving ejecta mass, and bound mass, respectively. Both the total ejecta and bound mass exhibit a strong correlation with spin. The mass of the fast-moving ejecta increases linearly up to $a_{\mathrm{BH}} = 0.6$, and grows exponentially beyond this point.
}
\end{figure*}

\begin{figure*}
\centering
\includegraphics[width=\textwidth]{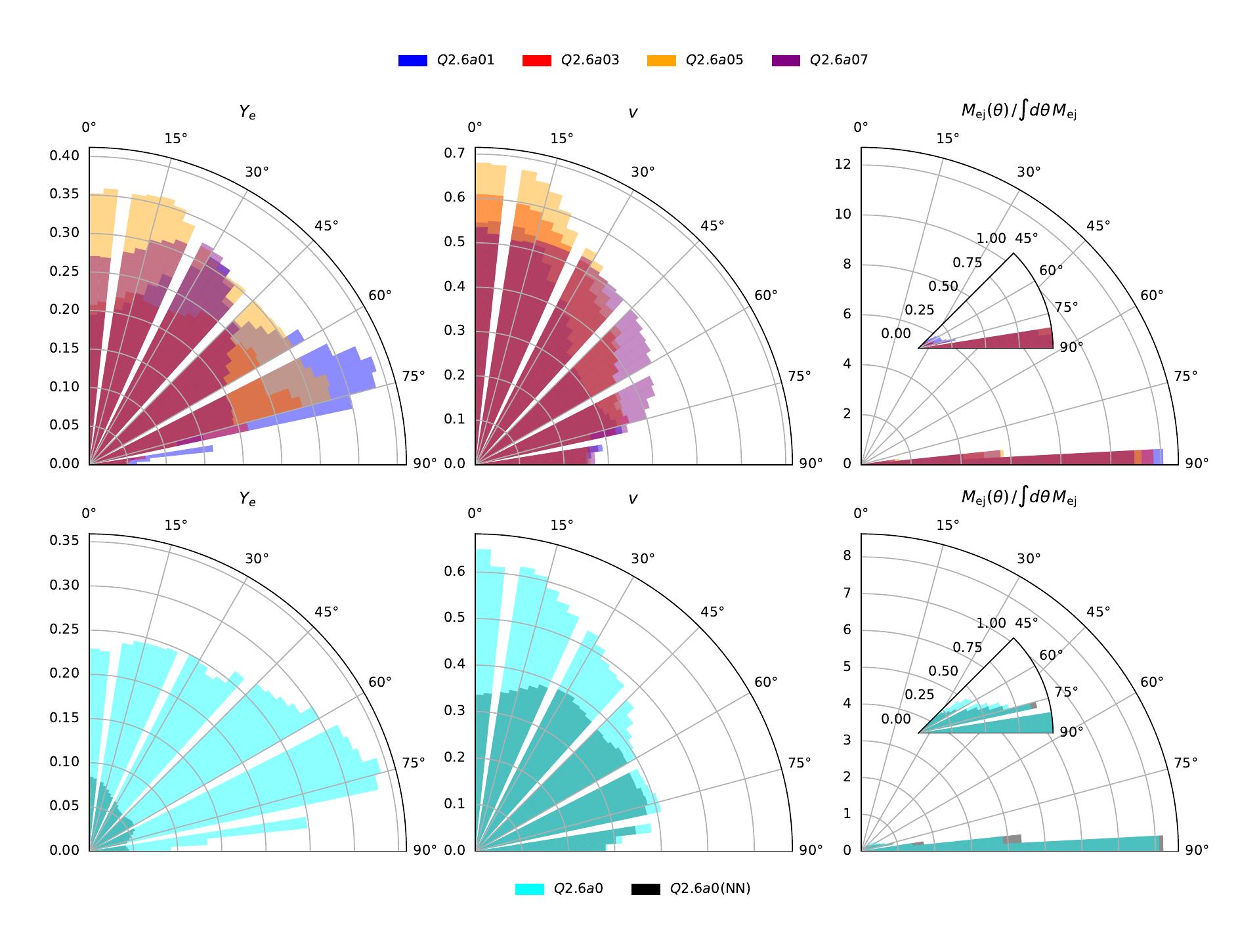}
\caption{\label{fig:angulardynamical} Angular distribution of the electron fraction, ejecta velocity, and normalised total dynamical ejecta mass at $\qty{7}{ms}$ after merger. The top panels compare different spin cases by showing the \texttt{Q2.6a01}, \texttt{Q2.6a03}, \texttt{Q2.6a05}, and \texttt{Q2.6a07} models, while the bottom panels present models with and without neutrino emission, comparing \texttt{Q2.6a0} and \texttt{Q2.6a0(NN)}. Most of the ejecta mass is concentrated near the equatorial plane, within only $10^\circ$. The figure also illustrates how the inclusion of neutrinos reprocesses the ejecta to higher $Y_{\mathrm{e}}$, particularly in the polar regions. The small gaps between bins are visual artefacts that appear when histogram bins are merged near the origin.}
\end{figure*}

In~\cref{fig:compsejecta200} we explore the relationship between \gls{BH} spin and the different components of the ejected matter. Additionally, the top panels of~\cref{fig:angulardynamical} illustrate the angular distributions of the electron fraction $Y_{\mathrm{e}}$, the ejecta velocity $v$, and the mass of dynamical ejecta, $M_{\mathrm{ej}}$.

As shown in both~\cref{table:finalproperties300} and~\cref{fig:compsejecta200}, the total dynamical ejecta, fast-moving ejecta, and bound mass increase with black hole spin. This trend reflects the fact that higher spin enhances tidal disruption, which drives more efficient mass ejection. Although the overall dependence is not strictly linear, the first six models display an approximately linear behavior. Beyond this point, the quantities grow more rapidly, indicating a transition to a nonlinear, possibly exponential regime. These trends are significant, as the ejecta mass plays a central role in shaping the outcome of $r$-process nucleosynthesis and the brightness of the associated kilonova~\citep{2013AAS...22134604B, 2016ApJ...825...52K}.

When focusing on the first six models, the relationship between spin and the total ejecta mass yields a coefficient of determination ($R^2$) of approximately $0.77$ (which indicates how the fit is consistent with the data), indicating a reasonably strong linear trend. In comparison, the disc mass shows an even stronger linear correlation with spin, with an $R^2$ value of $0.97$. This high degree of linearity makes the disc mass a particularly relevant quantity for interpreting the electromagnetic counterpart.

The most important exception is the \texttt{Q2.6a08} model, where both the dynamical ejecta and bound mass are lower than in the \texttt{Q2.6a06} model. As can be seen from~\cref{table:conv}, at lower resolution the total dynamical ejecta mass also follows this exponential trend, increasing by a factor of approximately $2.5$ compared to the \texttt{Q2.6a07} model. Therefore, any non-monotonic behaviour can be attributed to resolution effects, although a strictly monotonic trend is not necessarily expected.

We also examined the ejecta under two conditions: i) $Y_{\mathrm{e}} > 0.25$, and ii) $Y_{\mathrm{e}} \leq 0.25$. In this context, condition (i) corresponds to neutron-poor ejecta, while condition (ii) represents neutron-rich ejecta. In all models, we find that the mass of the neutron-rich component is at least ten times greater than that of the neutron-poor component. Our simulations suggest a dimmer kilonova counterpart, as expected from former studies (see~\cite{reviewfoucart}). Across all simulations, the neutron-poor ejecta mass ranges between $10^{-6}$ and $10^{-4} M_\odot$. However, we note that these values reflect only the dynamical ejecta and do not include contributions from wind-driven ejecta.

\begin{table*}
\caption{The properties of the different ejecta components measured from a radius of $\qty{300}{\km}$ at $\qty{7}{\ms}$ after the merger. The columns list the model name, total ejecta mass, mass-averaged electron fraction of the total ejecta, mass of the fast-moving ejecta, mass of the bound material, disc mass, the mass of the spiral wind ejecta at $\qty{3}{ms}$ after the merger, and the mass of the spiral wind ejecta while the change in ejecta is under $10$ per cent. The label (NN) denotes simulations performed without neutrino transport (no-neutrino). The masses of all ejecta components generally increase with spin, except in a few models. While the electron fraction is notably high in the non-spinning and \texttt{Q2.6a08} models compared to the other simulations, lower spin values tend to produce less neutron-rich ejecta overall. A significant amount of spiral wind ejecta is detected in all models.}\label{table:finalproperties300}
\small
\begin{tabular}{@{}lSSSSSSSSSS@{}}
\hline
	   Model &$\unit{\Mass}_ {\mathrm{ej}}  [\qty{e-2}{\Mass\Sun}]$& $Y_ {\mathrm{e}}$ & $\unit{\Mass}_ {\mathrm{fast}}  [\qty{e-5}{\Mass\Sun}]$ & $\unit{\Mass}_ {\mathrm{fb}}[\qty{e-2}{\Mass\Sun}]$ &$\unit{\Mass\disk} [\unit{\Mass\Sun}]$ &$\unit{\Mass}_ {\mathrm{sp,1}}[\qty{e-2}{\Mass\Sun}]$ &$\unit{\Mass}_ {\mathrm{sp,2}}[\qty{e-4}{\Mass\Sun}]$\\
		\hline
        \texttt{Q2.6a0}  & 0.020  &0.160   &1.570   &0.110    &0.028 & 0.021 & 0.42\\ 
        \texttt{Q2.6a0(NN)} & 0.019 & 0.046 & 0.612 & 0.110 & 0.028 & 0.013 & 3.31\\
        \texttt{Q2.6a01} & 0.056  &0.087   &0.047   &0.250    &0.056 & 0.029 & 1.00\\
        \texttt{Q2.6a02} & 0.047  &0.066   &0.280   &0.290    &0.061 & 0.104 & 1.82\\ 
        \texttt{Q2.6a03} & 0.120  &0.062   &1.190   &0.560    &0.104 & 0.152 & 2.46\\ 
        \texttt{Q2.6a04} & 0.296  &0.060   &2.250   &1.050    &0.129 & 0.102 & 5.80\\ 
        \texttt{Q2.6a05} & 0.589  &0.055   &1.290   &1.730    &0.151 & 0.276 & 12.20\\ 
        \texttt{Q2.6a06} & 1.230  &0.057   &9.070   &2.690    &0.110 & 0.556 & 29.10\\ 
        \texttt{Q2.6a07} & 2.050  &0.057   &26.150  &3.690    &0.143 & 1.652 & 47.40\\ 
        \texttt{Q2.6a08} & 0.780  &0.140   &88.630  &1.610    &0.153 & 2.373 & 72.30\\ 
        \texttt{Q2.6a08(NN)} & 3.38 & 0.050 & 27.700 & 3.801 & 0.226 & 2.863 & 194.70\\
	    \hline
\end{tabular}
\end{table*}

In~\cref{fig:angulardynamical}, we show the angular distribution of the electron fraction $Y_{\mathrm{e}}$, velocity $v$, and ejecta mass for the \texttt{Q2.6a01}, \texttt{Q2.6a03}, \texttt{Q2.6a05}, and \texttt{Q2.6a07} models, in the top panels. As shown in the figure, the neutron-poor component of the ejecta ($Y_{\mathrm{e}} \geq 0.25$) accumulates in the polar region ($0^\circ \leq \theta \leq 45^\circ$), while the equatorial region ($\theta \geq 80^\circ$) contains very neutron-rich material (with $0.05 \leq Y_{\mathrm{e}} \leq 0.2$). The only exception is the \texttt{Q2.6a02} model, which is not included in the figure. We note that, in the polar region, the \texttt{Q2.6a02} model produces more dynamical ejecta than the other models for spins up to 0.5. In addition, in this model the maximum temperature shows two distinct peaks, which may modify the ejecta composition.

In the middle panel, we observe that the ejecta velocity exceeds $0.5\,c$ in the polar region, especially for $0^\circ \leq \theta \leq 20^\circ$, and gradually decreases towards the equator, dropping below $0.3\,c$ beyond $\theta = 75^\circ$. However, as shown in the right panel, most of the ejected mass is concentrated in the equatorial region ($\theta \geq 75^\circ$), indicating that the bulk of the ejecta is both highly neutron-rich and relatively slow, moving at approximately $0.2$–$0.3\,c$.

A peculiar behaviour in the $Y_{\mathrm{e}}$ distribution appears around $\theta = 40^\circ$. For models with lower black hole spin, $Y_{\mathrm{e}}$ remains low up to this angle and increases beyond it. This suggests that in higher spinning configurations, a brighter electromagnetic counterpart could emerge from the polar regions, provided they contain a sufficient amount of ejecta.

We emphasize that the degree of tidal disruption increases with black hole spin. As clearly seen in~\cref{fig:angulardynamical}, more matter is ejected towards the equatorial plane, and in this region, the electron fraction $Y_{\mathrm{e}}$ decreases with spin. Although models that include neutrino transport exhibit higher $Y_{\mathrm{e}}$ overall, due to protonization (or leptonization)~\citep{reviewkyutoku} via the electron capture process ($\nu_e + n \rightleftharpoons p + e^-$), the observed trend of decreasing $Y_{\mathrm{e}}$ with spin may also be influenced by a similar interaction process. This can be examined by analyzing the neutrino luminosities and energies. Our results show that, the average luminosity and the total energy of heavy-lepton neutrinos dominate over all other neutrino species. The processes responsible for producing heavy-lepton neutrinos include pair processes, i.e., electron-positron annihilation ($e^+ + e^- \rightarrow \nu + \bar{\nu}$), plasmon decay ($\gamma + \gamma \rightarrow \nu + \bar{\nu}$), and nucleon-nucleon bremsstrahlung ($N + N \rightarrow \nu + \bar{\nu} + N + N$)~\citep{massejection, foucartneutrinoreview}. These interactions generate heavy-lepton neutrino-antineutrino pairs and therefore tend to reduce the electron fraction, $Y_{\mathrm e}$. 

As shown in~\cref{table:finalproperties300}, both models $Q2.6a06$ and $Q2.6a07$ exhibit the lowest $Y_{\mathrm e}$ together with the highest total heavy-lepton neutrino luminosities. This correlation suggests that enhanced pair processes at higher spin might be responsible for the reduced $Y_{\mathrm e}$ observed in these cases.

We also estimate the remnant baryonic mass using the fitting formula of~\cite{diskFoucart}. As described in~\cref{sec:numericalmethods}, when post-processing the disc masses we map our domain onto a uniform grid. On this grid, we compute the mass at both low resolution ($\qty{1}{\kilo\metre}$) and high resolution ($\qty{221}{\metre}$) over a region extending to $\qty{300}{\kilo\metre}$. The difference between these resolutions affects the disc mass by at most a few percent for spins up to $a_{\mathrm{BH}} = 0.4$, but the magnitude of this change increases with spin. Following the method of~\cite{diskFoucart}, we then calculate the remnant baryonic mass and find good agreement with our results up to $a_{\mathrm{BH}} = 0.6$, except for the \texttt{Q2.6a0} model, which has a very low disc mass. If we exclude the outlier model \texttt{Q2.6a08}, our disc mass is about $36$ per cent lower than the estimate from~\cite{diskFoucart}. However, when examining the convergence test for this model, the relative error of the disc is $32$ per cent.

The relative error for the \texttt{Q2.6a0} model is negligible for the MR and LR resolutions. As mentioned, the \texttt{Q2.6a08} model yields a higher relative error of $32$ per cent, which is consistent with the results reported in~\cite{massejection}.

\begin{figure*}
\centering
\includegraphics[width=0.5\textwidth]{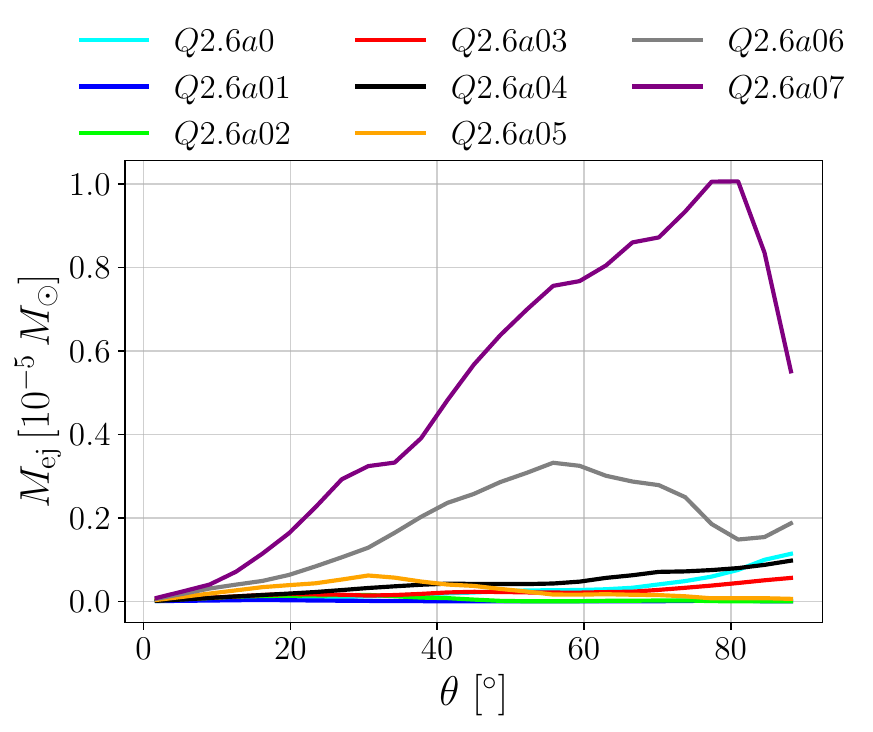}
\caption{\label{fig:fast} The angular distribution of the fast-moving ejecta for all cases without NN models is shown. As seen, most of the ejecta is confined to the equatorial plane, particularly at angles $\theta > 30^\circ$. This angular confinement indicates that this component corresponds to the sprayed-out fast-moving ejecta.}
\end{figure*}

\begin{table*}
\caption{The convergence tests results for \texttt{Q2.6a0} and  \texttt{Q2.6a08} model. The columns list the model name, total ejecta mass, mass-averaged electron fraction of the total ejecta, mass of the fast-moving ejecta, mass of the bound material, disc mass, the mass of the spiral wind ejecta at $\qty{3}{ms}$ after the merger, and the mass of the spiral wind ejecta while the change in ejecta is under $10$ per cent. We do not report the spiral wind-driven ejecta for the \texttt{Q2.6a08} model, since the dynamical mass ejection does not reach saturation even after $\qty{12}{\ms}$.   }\label{table:conv}
\small
\begin{tabular}{@{}lSSSSSSSSSS@{}}
\hline
	   Model &$\unit{\Mass}_ {\mathrm{ej}}  [\qty{e-2}{\Mass\Sun}]$& $Y_ {\mathrm{e}}$ & $\unit{\Mass}_ {\mathrm{fast}}  [\qty{e-5}{\Mass\Sun}]$ & $\unit{\Mass}_ {\mathrm{fb}}[\qty{e-2}{\Mass\Sun}]$ &$\unit{\Mass\disk} [\unit{\Mass\Sun}]$ &$\unit{\Mass}_ {\mathrm{sp,1}}[\qty{e-2}{\Mass\Sun}]$ &$\unit{\Mass}_ {\mathrm{sp,2}}[\qty{e-4}{\Mass\Sun}]$\\
		\hline
        \texttt{Q2.6a0$_{MR}$}  & 0.029  &0.153   &0.544   & 0.135   &0.027 & 0.015 & 0.69\\ 
        \texttt{Q2.6a0$_{LR}$} & 0.039 & 0.166 & 0.286 & 0.142 & 0.025 & 0.017 & 0.77\\
        \texttt{Q2.6a08$_{MR}$} & 5.186  &0.140   & 22.60  &1.204    &0.103 & NA & NA \\
	       \hline
\end{tabular}
\end{table*}

We define the ejecta as fast-moving if its velocity exceeds $0.6\,c$. This component is particularly important, as its high velocity allows it to interact with the surrounding environment and produce an additional \gls{EM} counterpart~\citep{2015MNRAS.446.1115M, Radicefast}. To investigate this effect, we analyze the properties of fast-moving ejecta and their dependence on the initial \gls{BH} spin, as shown in~\cref{fig:compsejecta200}. Up to the \texttt{Q2.6a06} model, the fast-moving ejecta exhibits an approximately linear trend with spin. Beyond this point, the dependence becomes clearly nonlinear, with a more quadratic character.

We further examine the angular distribution of fast-moving ejecta. For all models, the electron fraction $Y_{\mathrm{e}}$ of the fast-moving ejecta remains below $0.4$ in the polar region ($\theta \leq 40^\circ$) and spans a broader range ($0.1 \leq Y_{\mathrm{e}} \leq 0.8$) at higher latitudes. Given the very low mass in the polar region, the unusually high $Y_{e}$ values may be numerical artifacts.
In the region $\theta \geq 40^\circ$, its $Y_{\mathrm{e}}$ generally decreases with increasing spin. The maximum velocity of the fast-moving ejecta reaches up to $0.85\,c$, particularly in the polar regions.

The mass distribution of fast-moving ejecta is nearly uniform across polar and equatorial directions, for spins up to $0.6$. For higher spins, it becomes increasingly concentrated toward the equatorial plane. 
Notably, the \texttt{Q2.6a02} model shows no fast-moving material within the angular range $40^\circ \leq \theta \leq 60^\circ$.

The angular distribution and electron fraction analysis suggest that a blue \gls{EM} component may emerge from the fast-moving ejecta in the equatorial region.

Earlier work by~\cite{most2020tab} reported that the mass of the fast-moving ejecta is negligible, reaching at most $\qty{1.4e-8}{\Mass\Sun}$ for a system with $M_{\mathrm{BH}} = \qty{2.42}{\Mass\Sun}$, $M_{\mathrm{NS}} = \qty{1.18}{\Mass\Sun}$, and $a_{\mathrm{BH}} = 0.52$. In contrast, in our earlier study~\citep{matur2024tab}, we found evidence for fast-moving ejecta even though the mass ratio was higher. In the present work, we confirm that result. Our findings show that, even with a slightly higher mass ratio and a non-spinning \gls{BH}, the fast-moving ejecta can reach $\qty{1.57e-5}{\Mass\Sun}$, which is comparable to their BNS merger results, despite our definition of fast-moving ejecta adopting a stricter velocity threshold ($v \geq 0.6\,c$).

In \gls{BNS} mergers, fast-moving ejecta are expected even in binaries that lead to prompt \gls{BH} formation, as shown in \cite{massejection}, \cite{2026MNRAS.545f2009K}, and \cite{2025MNRAS.538..907R}. In the latter study, which uses Smoothed Particle Hydrodynamics (SPH) simulations allowing for accurate tracking of particle trajectories, the fast-moving ejecta are divided into two categories: \textit{sprayed-out} and \textit{bounced}. By following the classification proposed in \cite{2025MNRAS.538..907R}, we characterise the sprayed-out fast-moving ejecta in our simulations using the following properties:
\begin{enumerate}
    \item Angular distribution: predominant confinement to a narrow region around the orbital plane,
    \item Remnant oscillations: Even though the prompt-collapse cases studied in \citet{2025MNRAS.538..907R} exhibit both spray-out and bounce-back fast-moving ejecta, the presence of two neutron stars during the merger phase allows part of the material to be temporarily confined near the central region. In our case, however, we do not identify any physical mechanism that could lead to a subsequent bounce-back of this centrally confined material following the merger.
\end{enumerate}
As our results satisfy these conditions (see the velocity and mass distributions in \cref{fig:angulardynamical} and the angular distribution of the fast-moving ejecta in \cref{fig:fast}), we classify the fast-moving ejecta as sprayed-out. We also note that our criterion for defining fast-moving ejecta is $v \geq 0.6\,c$, which is higher than the threshold adopted in \cite{2025MNRAS.538..907R}, where fast-moving ejecta are defined as having $v > 0.4\,c$. Therefore, while a fast component is identified in all cases, our simulations do not yet demonstrate that its mass converges to a finite value, and its interpretation as a sprayed-out component requires further justification.

\begin{figure*}
\centering
\includegraphics[width=\textwidth]{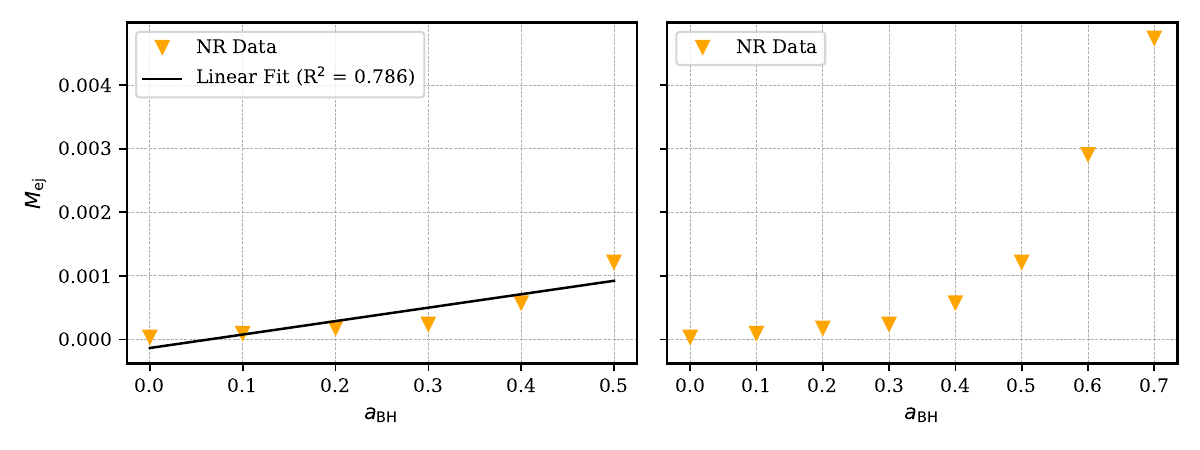}
\caption{\label{fig:disk}The relationship between the spiral wind-driven ejecta and spin. The left and right panels show the variation of the spiral wind-driven ejecta with spin in the ranges \numrange[range-phrase=--]{0}{0.5} and \numrange[range-phrase=--]{0}{0.7}, respectively. While the lower-spin models exhibit a roughly linear correlation between spin and ejecta mass, the overall trend remains consistent across most of the spin range.}
\end{figure*}

\begin{figure*}
\centering
\includegraphics[width=\linewidth]{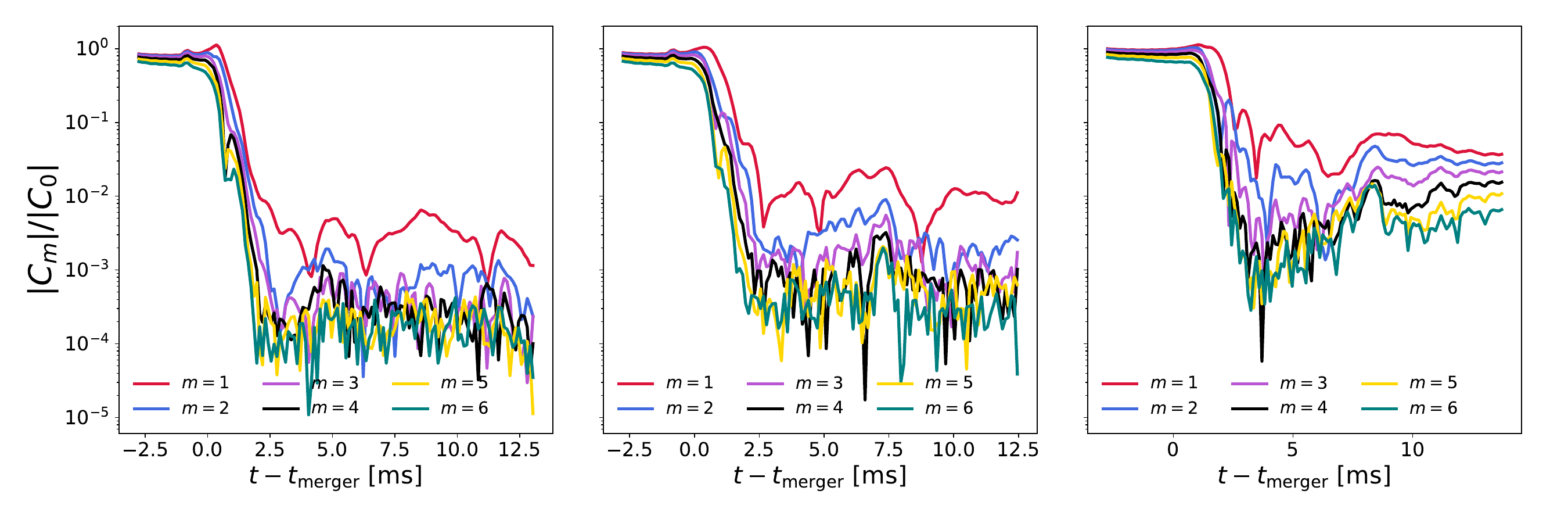}
\caption{\label{fig:mode_amplitudes}The evolution of the density modes for \texttt{Q2.6a01}, \texttt{Q2.6a04} and \texttt{Q2.6a07} models. The extraction radius for the modes is $\qty{300}{\km}$. Different colours represent different modes. The $x$-axis indicates the time elapsed after the merger, while the $y$-axis shows the mode amplitude, normalized by the $m = 0$ mode at merger. The $m = 1$ mode is dominant, and its amplitude increases with the dimensionless spin parameter of the initial black hole. The amplitude grows by a factor of $\num{\sim 4}$ between \texttt{Q2.6a04} and \texttt{Q2.6a07}, and by a factor of $\num{\sim 10}$ between \texttt{Q2.6a0} and \texttt{Q2.6a07}.
}
\end{figure*}

\subsection{Spiral wind and density oscillations}

As defined in Section~\ref{sec:numericalmethods}, in this work, following previous studies~\citep{spiral1, spiral2}, we define the spiral-wind ejecta as all late-time outflow emerging after the saturation of the dynamical ejecta. In our simulations, we find that the dynamical ejecta saturates significantly faster than in \gls{BNS} mergers, which is expected since there is no central hypermassive neutron star to keep feeding the ejecta. 

While the dynamical ejecta grow exponentially within just $\sim \qty{3}{ms}$ after the merger in all cases, we calculate them both from this early time ($M_{\mathrm{sp,1}}$) and later when the total ejecta mass change falls below $10$ per cent ($M_{\mathrm{sp,2}}$). The spiral wind ejecta mass lies in the range $2.1 \times 10^{-3} \leq M_{\mathrm{sp,1}} \leq 1.652 \times 10^{-2}~M_\odot$ and in the range $4.21 \times 10^{-5} \leq M_{\mathrm{sp,2}} \leq 4.74 \times 10^{-3}~M_\odot$ when excluding the \texttt{Q2.6a08} model. In both cases, despite a reduction by factors of $3-5$ between maximum and minimum values, the spiral wind ejecta remains present.

Here, we emphasise that the spiral wind-driven ejecta does not represent the entire post-merger ejecta, nor does it encompass all non-dynamical ejecta, which includes, in particular, the shock-heating-driven component. In the literature, different mass-ejection mechanisms are commonly identified and quantified using distinct criteria (e.g.~\cite{2021PhRvD.104l3010F}). Specifically, the dynamical ejecta is identified using the geodesic criterion ~\citep{massejection, spiral1}, whereas the shock-heating-driven ejecta is calculated based on the Bernoulli criterion and saturates very fast~\citep{2015MNRAS.450.1430H}. The neutrino-driven ejecta is restricted by its angular distribution, being confined to the polar regions~\citep{massejection}. The spiral wind-driven ejecta constitutes a subset of the post-merger non-dynamical ejecta. Unlike the other components, however, it is not measured immediately after merger. Instead, as explained in the previous paragraph, it is quantified only after the dynamical mass ejection has saturated, by considering a specific post-merger time interval. This selection prevents contamination from the other parts of the ejecta and allows the spiral wind-driven component to be isolated. However, we caution that this component represents a subset of the non-dynamical ejecta as identified by the Bernoulli criterion.

Since our simulations continue to about $\qty{13}{ms}$ post-merger, this provides a $\qty{10}{ms}$ window to analyse the subsequent spiral wind-driven outflow. The mass of the spiral wind ejecta from all simulations is shown in~\cref{table:finalproperties300}. As shown in the table, the ejecta mass generally increases with the dimensionless spin, except for in the case of the \texttt{Q2.6a04} model. Moreover, the mass of the spiral wind-driven ejecta increases with increasing disc mass, as seen in~\citet{spiral1}.

According to~\cite{spiral1}, the spiral wind ejecta contributes to the blue, day-long kilonova emission. 
Therefore, our study suggests that the spiral wind in \gls{BHNS} mergers may contribute to a blue kilonova component. However, to confirm this interpretation, we still need to compute kilonova light curves.

Given that this ejecta component can significantly alter the electromagnetic signature of \gls{BHNS} mergers, we also investigate its physical origin. To this end, we perform an analysis similar to that of~\cite{eastspiral, spiral2} and present it in~\cref{fig:mode_amplitudes}. In our case, the only contribution to this behaviour comes from the remnant disc. According to~\cite{spiral2}, the odd modes are excited in the remnant hypermassive neutron star due to the low-$T/|{W}|$ instability. 

In~\cref{fig:mode_amplitudes}, we present the evolution of the density modes from $m = 1$ to $m = 6$ for the \texttt{Q2.6a01}, \texttt{Q2.6a04}, and \texttt{Q2.6a07} models, following~\cite{spiral2, eastspiral}. Compared to~\cite{spiral2}, we observe that the odd modes appear from the beginning of the simulation. The normalised mode amplitudes are comparable for all spin values, in the sense that the relative ordering of the different modes remains similar for each case. However, the high-spin models show systematically higher amplitudes, especially for the $m = 1$ mode. 
In all cases, the $m = 1$ mode clearly emerges as the dominant contribution. 

When examining the spiral wind ejecta across individual models, we find that its mass (in both $M_{\mathrm{sp,1}}$ $M_{\mathrm{sp,2}}$) increases by a factor of approximately $\sim47$ from the non-spinning case (\texttt{Q2.6a01}) to the \texttt{Q2.6a07} model, and by a factor of $\num{\sim 8}$ between the \texttt{Q2.6a04} and \texttt{Q2.6a07} models. The corresponding density mode amplitudes exhibit a similar trend, increasing by factors of $\num{\sim 30}$ and $\sim 4$, respectively, over the same model comparisons. This parallel scaling suggests that the spiral wind ejecta originates from disc oscillations. 

In the spiral wind-driven ejecta, the convergence study shows that while the relative error in \texttt{Q2.6a0} for 
$M_{\mathrm{sp,1}}$  is around $30$ and $20$ per cent for MR and LR simulations, respectively, the error in  $M_{\mathrm{sp,2}}$ is much higher, as expected~\citep{spiral2}. We also find that the dynamical mass ejection for \texttt{Q2.6a08} model does not reach saturation even after $\qty{12}{ms}$. Therefore, to accurately determine the corresponding ejecta, high-resolution and longer-term simulations are required, which we leave for future work.

While this study provides the first demonstration of density mode oscillations in \gls{BHNS} mergers, similar analyses have previously been carried out in self-gravitating tori around spinning black holes~\citep{bhspiral}. Additionally,~\cite{eastdarkmatter} investigated the evolution of density modes in a neutron star gradually consumed by a central black hole, which may be conceptually related to the present scenario.

\begin{figure*}
\centering
\includegraphics[width=\textwidth]{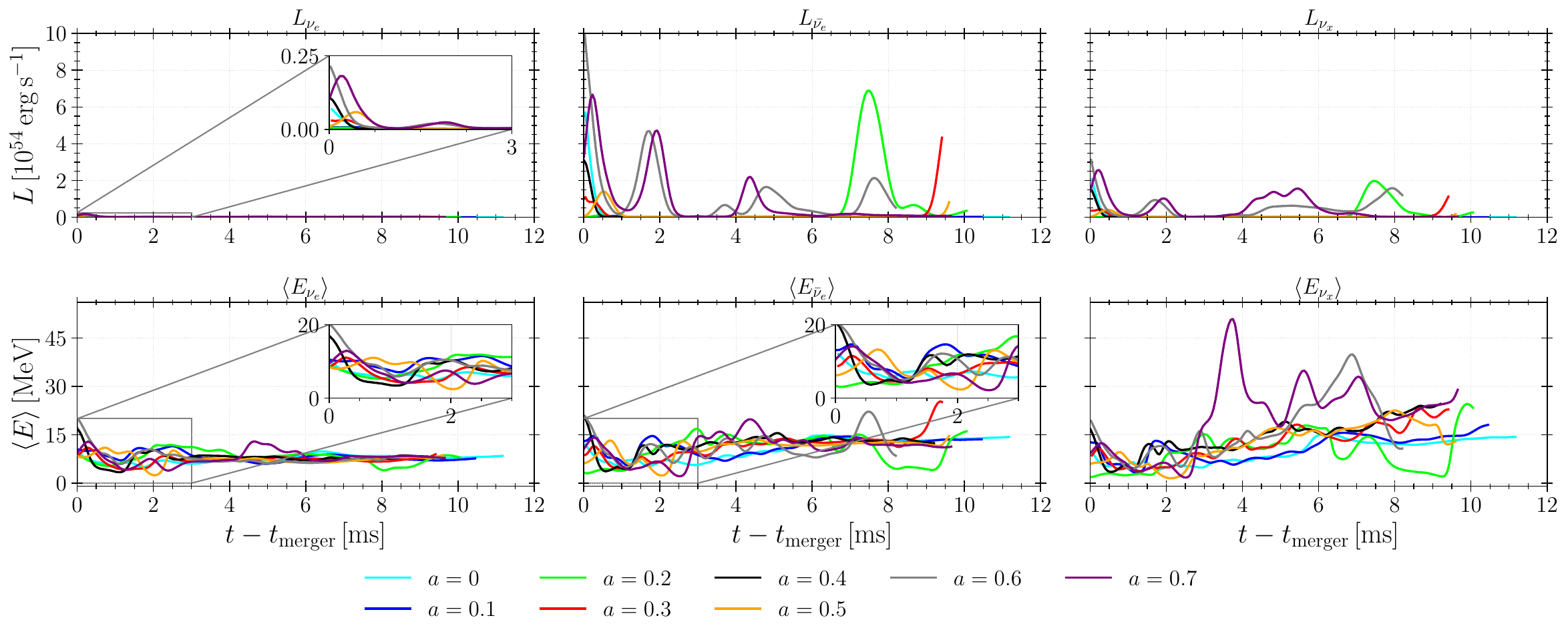}
\caption{\label{fig:neutrino} The neutrino luminosity and average energy for all models and for the three different neutrino flavours: electron neutrinos, electron antineutrinos, and heavy-lepton neutrinos. Both the luminosity and energy increase from electron neutrinos to electron antineutrinos and heavy-lepton neutrinos. Moreover, we observe a correlation between these quantities and the initial spin of the black hole. This trend suggests that neutrino properties could serve as an additional tool for constraining the initial black hole spin.}
\end{figure*}

\subsection{Neutrino effects}

In mergers involving neutron stars (either \gls{BNS} or \gls{BHNS} mergers), neutrino transport is particularly important because it alters the temperature, composition and distribution of the ejecta~\citep{massejection, duezneutrino}. Since the composition is the most critical property for $r$-process nucleosynthesis, it must be accurately determined to achieve reliable outcomes.

We first examine the role of neutrino effects on the composition of the ejecta by comparing the models \texttt{Q2.6a0}, \texttt{Q2.6a0(NN)}, \texttt{Q2.6a08}, and \texttt{Q2.6a08(NN)}. Here, the label \texttt{(NN)} denotes simulations performed without neutrino transport (no-neutrino). Although the highest mass-weighted electron fractions for the dynamical ejecta are observed in the non-spinning and the highest spin simulations -- specifically, \texttt{Q2.6a0} and \texttt{Q2.6a08} -- with values around $0.16$ and $0.14$, the no-neutrino simulations produce significantly more neutron-rich ejecta, with $Y_{\mathrm{e}} \approx 0.046$ and $0.05$, respectively. This indicates that neutrino interactions substantially change the composition of the ejecta, as expected, by increasing the electron fraction due to protonization of the matter. Consequently, simulations that include neutrino transport are expected to yield less favourable conditions for heavy $r$-process nucleosynthesis.

We further find that models without neutrino transport produce a narrow $Y_{\mathrm{e}}$ distribution, concentrated primarily in the \numrange[range-phrase = --]{0.0}{0.1} range,  extending only up to $0.4$. In contrast, models including neutrinos exhibit a broader, more uniform distribution reaching up to $Y_{\mathrm{e}} = 0.5$. This broader range implies the possibility of a wider variety of \gls{EM} counterparts, including both dim and bright kilonovae, rather than just faint transients due to high opacities.

In \gls{BHNS} mergers, neutrinos play a crucial role because they also heat the outflow. Since temperature evolution is one of the key factors for $r$-process nucleosynthesis, it is important to compare the temperature distributions for the neutrino and NN models. When we examine the evolution of the maximum temperature, we find that the matter is hotter in the simulations that include neutrino effects. This heating is due to neutrino irradiation, which raises the temperature of the ejecta, and in turn reprocess it to higher $Y_{\mathrm{e}}$.

Before discussing the neutrino energy and luminosities, we first focus on the (NN) values in~\cref{table:finalproperties300}. As shown in this table, while the total mass of the dynamical ejecta does not change much with neutrino transport in the non-spinning cases, we observe an increase in the (NN) model with a spin of $0.8$. As mentioned earlier, the mass of the ejected matter changes significantly with resolution. While the ejecta mass in the NN simulation is nearly the same as that in the MR simulation for the \texttt{Q2.6a08} model, the difference becomes large in the HR simulation. However, the most converged quantity, the mass-weighted electron fraction, is considerably higher compared to the NN model. In the non-spinning models, the masses of the fast-moving ejecta, fallback material, and the disc remain nearly unchanged, with the only notable difference coming from the spiralwind ejecta. 

Figure~\ref{fig:neutrino} shows the evolution of neutrino luminosities and average energies for three flavors: $\nu_e$, $\bar{\nu}_e$, and $\nu_x$. As shown, $\nu_e$ has the lowest luminosity, while $\nu_x$ reaches the highest values. A similar trend appears in the average energies, following $E_{\nu_e} < E_{\bar{\nu}_e} < E_{\nu_x} $, although the energy differences are not as large as the luminosity differences. However, although most of the energy comes from $\nu_x$, this group actually consists of four species: muon and tau neutrinos and anti-neutrinos. Accordingly, the luminosity and energy associated with each species are $E_{\nu_x}/4$ and $L_{\nu_x}/4$, respectively.

Among the different spin models, the \texttt{Q2.6a06} and \texttt{Q2.6a07} configurations show the highest peak luminosity. Interestingly, the luminosity decreases with decreasing spin, with the exception of the non-spinning model, which ranks fourth in terms of peak luminosity. Regarding neutrino energies, again, both $ E_{\nu_e} $, $ E_{\bar{\nu}_e}$ and $ E_{\nu_x}$ peak in the \texttt{Q2.6a06} and \texttt{Q2.6a07} models. This further demonstrates that spin not only affects the neutrino luminosity but also significantly impacts the spectral properties of emitted neutrinos.

According to the study by~\cite{foucart2015tab}, which employed the \texttt{M1} scheme for neutrino transport, simulations with $M_{\mathrm{BH}} = \qty{7}{\Mass\Sun}$, $M_{\mathrm{NS}} = \qty{1.4}{\Mass\Sun}$, and $a_{\mathrm{BH}} = 0.8$ found that the neutrino luminosities for electron neutrinos and anti-neutrinos peak at $\qty{e54}{\erg\per\second}$, in agreement with our results, while those for heavy-lepton neutrinos are the lowest, at $\qty{e53}{\erg\per\second}$. In contrast, we find that the heavy-lepton neutrino luminosities exceed those of the electron neutrinos. Specifically, our simulations yield heavy-lepton neutrino luminosities on the order of $\qty{e54}{\erg\per\second}$ for all models.

\cite{kyutoku2016tab} investigated the effects of neutrinos on \gls{BHNS} mergers with a mass ratio of $Q = 4$ and a black hole spin of $a_{\mathrm{BH}} = 0.7$. They found that neutrino irradiation does not significantly alter $Y_{\mathrm{e}}$. They also noted that, although the ejecta mass is quite small in the polar region, the $Y_{\mathrm{e}}$ is relatively high. Our results are consistent with their findings.

\begin{figure}
\centering
\includegraphics[width=0.5\textwidth]{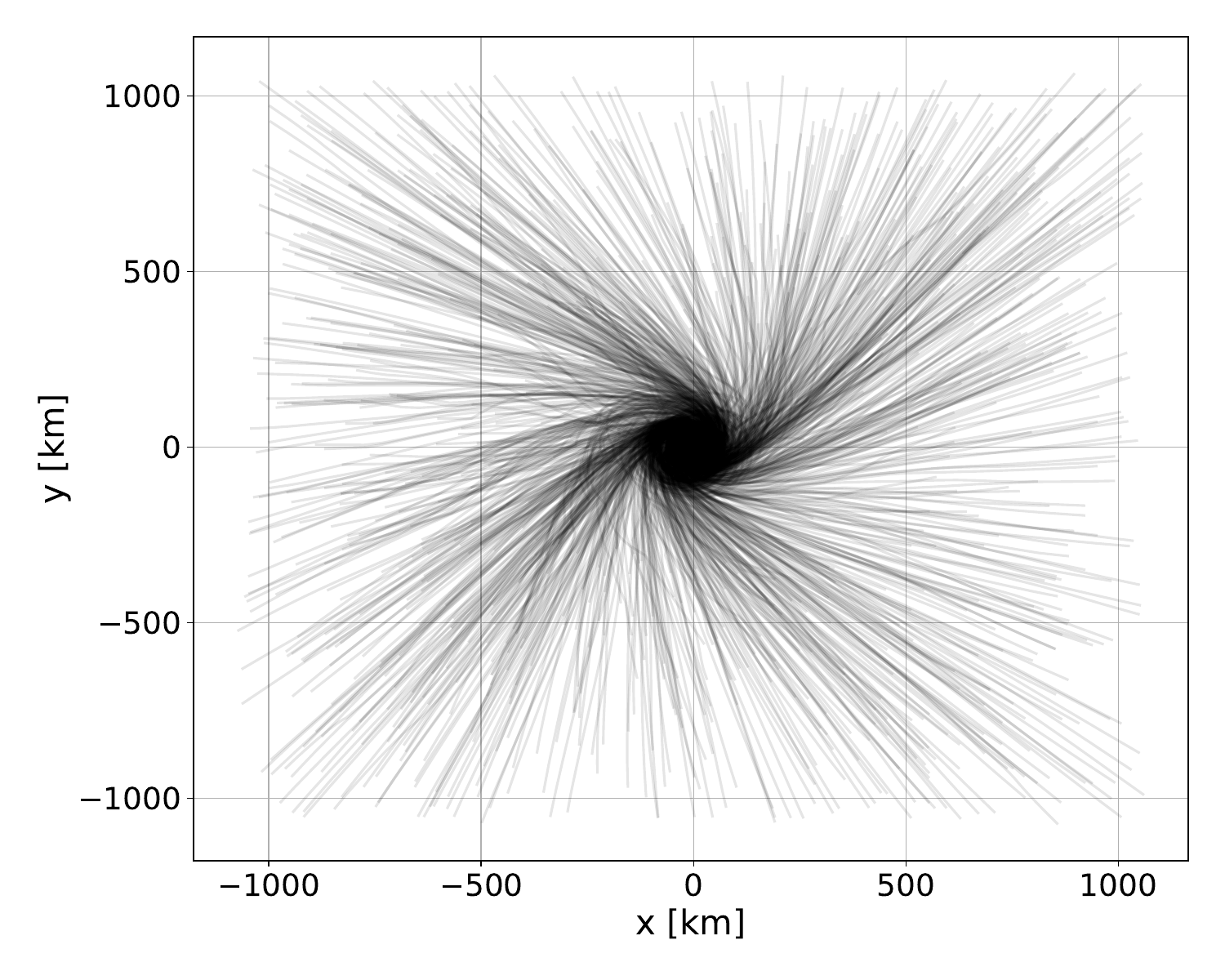}
\caption{\label{fig:trajectories} Trajectories of $1000$ tracer particles for the \texttt{Q2.6a0} model. After identifying the positions of the ejected matter, we compute the hydrodynamical properties of the corresponding tracers to provide input for $r$-process nucleosynthesis calculations. The particles are initially located inside the neutron star and are then ejected after the merger. We use $40$ CPU cores and generate the tracers within approximately $3$ hours per simulation.}
\end{figure}

\begin{figure}
\centering
\includegraphics[width=0.5\textwidth]{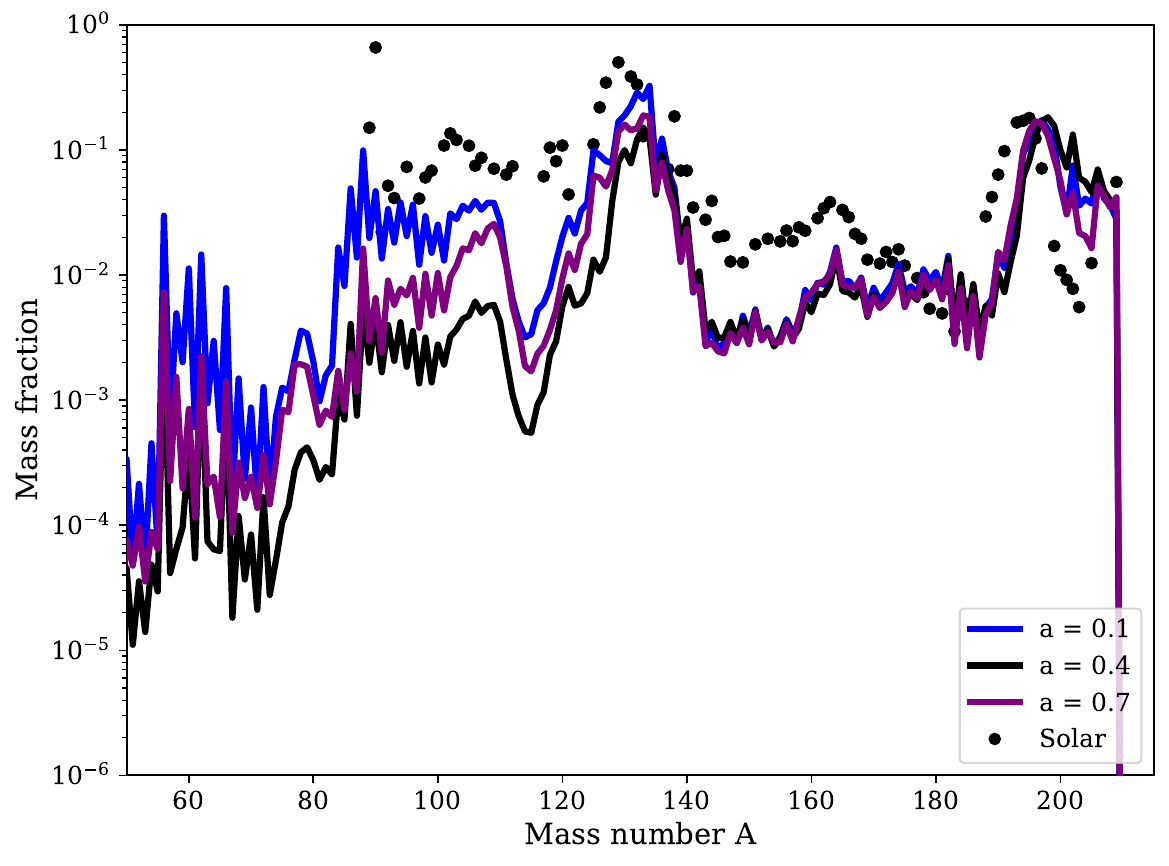}
\caption{Normalized $r$-process nucleosynthesis results for the \texttt{Q2.6a01}, \texttt{Q2.6a04}, and \texttt{Q2.6a07} models. Normalisation is performed by fixing the total mass fraction of elements with $180 \leq A \leq 200$. While elements $A \leq 120$ show some distinguishable differences, the heavier elements exhibit similar trends, as expected from their $Y_{\mathrm{e}}$ distributions. Although certain mass ranges are underproduced, the overall abundance patterns are generally consistent with the solar abundances.}\label{fig:nucleo}
\end{figure}

\subsection{$r$-process nucleosynthesis}\label{sec:rprocess}

Whilst discussing the properties of the ejected matter, we note that the dynamical ejecta is very neutron-rich, and therefore, heavy $r$-process nucleosynthesis is expected. However, we need to perform more detailed analysis to make more confident statements. For this purpose, we carry out the $r$-process nucleosynthesis for the dynamical ejecta as described in~\cref{sec:numericalmethods} using \texttt{WinNet}. The results for three simulations -- namely, \texttt{Q2.6a01}, \texttt{Q2.6a04}, and \texttt{Q2.6a07} spin models -- are shown in~\cref{fig:nucleo}. In this figure, we present the element abundances for mass numbers in the range of $50 \leq A \leq 215$ for these models. We choose them to illustrate the differences observed in some spinning \gls{BHNS} merger simulations. The solar abundances are taken from~\cite{1999ApJ...525..886A} for comparison.

Initially, we performed the nucleosynthesis calculation using only four tracer particles per model. When increasing the number to twenty, we already observe significant differences in the results. Therefore, we decide to calculate $1000$ tracers to ensure that the $r$-process nucleosynthesis results are independent of the choice of ejected particles. An example is shown in~\cref{fig:trajectories}, which displays the tracer trajectories for the \texttt{Q2.6a0} model. As shown, all tracer paths can be traced back to the initial position of the \gls{NS}, confirming their dynamical origin.

We can divide~\cref{fig:nucleo}, which shows the abundances normalised by fixing the total mass fraction of elements with $180 \leq A \leq 200$ (following~\cite{massejection}), into two regions: one from $A = 60$ to $120$, and the other from $A = 120$ to $215$. As shown in the figure, the highest abundances occur in the \texttt{Q2.6a01} model, which corresponds to the lowest-spin case in the figure, while the lowest abundances are seen in the \texttt{Q2.6a04} model. For lighter elements, the difference between \texttt{Q2.6a01} and \texttt{Q2.6a04} reaches nearly two orders of magnitude. However, since the corresponding mass ejection for these light-element abundances is small, the resulting ordering should be interpreted with caution, and further convergence tests are required to assess this trend more reliably. 

Despite this, \gls{BHNS} mergers produce fewer light $r$-process elements ($A < 120$) compared to BNS mergers (see the $r$-process results in~\cite{massejection}). For heavier elements ($A > 120$), the abundances increase, and all models produce them in comparable amounts. The differences between models are minor in this range, which is consistent with the $Y_{\mathrm{e}}$ distributions of the dynamical ejecta (see~\cref{fig:compsejecta200}). 

To compare the solar abundances with our results, it is conventional to examine the first, second, and third $r$-process peaks relative to the solar pattern~\citep{massejection, RosswogEM1, robertsnucleo, kyutoku2016tab}. According to this comparison, our results are generally consistent with the solar abundances, however, certain elements are underproduced due to the low $Y_{\mathrm{e}}$. Because $Y_{\mathrm{e}}$ is very low, we do not expect \gls{BHNS} mergers to produce elements below the second peak~\citep{kyutoku2016tab}.

In the study of~\cite{robertsnucleo}, they consider a $\sim 6$ mass-ratio \gls{BHNS} system with $M_{\mathrm{BH}} = \qty{7}{\Mass\Sun}$, $M_{\mathrm{NS}} = \qty{1.2}{\Mass\Sun}$, and $a_{\mathrm{BH}} = 0.9$. Although the mass ratio in their setup is quite large compared to our configurations, the high spin of the \gls{BH} is expected to produce a non-negligible amount of ejecta. This ejecta undergoes $r$-process nucleosynthesis, and because the $Y_{\mathrm{e}}$ is very low, it leads to the production of heavy elements. Their results show that the mass of neutron-rich ejecta ($Y_{\mathrm{e}} = 0.05$) is about $\qty{0.08}{\Mass\Sun}$, which is comparable to our \texttt{Q2.6a07} model, though our model yields about four times more ejecta with a similar composition. Figure 6 of~\cite{robertsnucleo} presents results very similar to ours. For the same configuration under a fixed neutrino irradiation assumption, they also found that elements with higher mass numbers follow a similar trend, while only those with mass numbers below $\sim 100$–$130$ show noticeable differences. We also find consistent results, not when changing the neutrino radiation, but when keeping the configuration fixed and varying the spin.

\section{Conclusion}\label{sec:conclusion}

Although black hole–neutron star mergers involving spinning black holes have been extensively studied in the literature, systematic investigations focusing on low mass-ratio systems remain limited. This work addresses that gap by presenting a suite of $11$ simulations that explore $9$ different dimensionless spin parameters ($a_{\mathrm{BH}} = 0.0$ to $0.8$ in steps of $0.1$), all aligned with the orbital angular momentum. Two configurations ($a_{\mathrm{BH}} = 0.0$ and $0.8$) are simulated both with and without neutrino transport, allowing us to isolate the effects of spin as well as neutrino interactions. The binary system modeled in this study lies within the chirp mass range consistent with GW230529, making the analysis directly astrophysically relevant. Our primary objective is to investigate how key observables are impacted by the black hole spin.

As shown in~\citet{campanellispin, etienne2009}, the spin–orbit interaction acts as a repulsive force against gravity for aligned spins. In line with this, we observe that increasing the black hole spin delays the merger for systems. We also see that all systems experience tidal disruption, but higher spin values lead to more violent mergers, leaving more baryonic matter outside the black hole and resulting in higher density and temperature in the remnant.

Performing simulations across a range of black hole spin values, we investigate how the spin affects the properties of the ejected matter. We find strong correlations, particularly in the amount of bound matter, fast-moving ejecta, spiral wind ejecta, and disc mass. These correlations are especially clear for the moderate spin range of $0 \leq a \leq 0.5$. Our findings suggest that electromagnetic counterparts, such as kilonovae, could potentially be used to constrain the spin parameter of the black hole. However, a direct kilonova light-curve analysis is required to confirm this possibility. 

Additionally, the dense and hot environments formed in these mergers provide ideal conditions for physical processes such as heavy element production. We observe that the electron fraction remains very low, especially for spin values in the range $0.1 \leq a \leq 0.7$, which is expected to produce heavy $r$-process elements.

In our simulations, the highest-spin models (\texttt{Q2.6a06} and \texttt{Q2.6a07}) yield the maximum luminosity and average energy for all neutrino flavours. We also observe clear evidence of neutrino heating in the outflows of models with neutrino transport, due to neutrino irradiation. This heating not only raises the temperature of the ejecta, but also alters its composition, resulting in a broader distribution of electron fractions.

In contrast, simulations without neutrino treatment yield a much narrower and more neutron-rich distribution ($Y_{\mathrm{e}} \lesssim 0.1$), which typically leads to dim kilonovae. However, the presence of neutrinos introduces a wider $Y_{\mathrm{e}}$ range -- up to $Y_{\mathrm{e}} \sim 0.5$ -- making bright electromagnetic counterparts possible. This demonstrates the essential role of neutrinos in shaping both the nucleosynthesis outcomes and observable signatures of black hole–neutron star mergers.

The most prominent results arise from the spiral wind–driven ejecta. While oscillations of the hypermassive neutron star are known to generate spiral winds that can drive outflows in binary neutron star mergers~\citep{spiral1}, this type of ejecta has not previously been reported for black hole–neutron star mergers. Our findings suggest that such ejecta are likely present in these systems. Spiral wind ejecta are expected to produce a blue, day-long kilonova. Although we do not perform a detailed kilonova light-curve analysis here, our results indicate the possibility of a blue kilonova component arising from black hole–neutron star mergers.

As a follow-up study, we plan to focus on kilonova light curve analysis. In particular, we aim to demonstrate the potential for a blue kilonova counterpart in black hole–neutron star mergers.

Although our simulations are performed with a standard resolution of $\qty{221}{\m}$, a higher resolution may help clarify certain features, especially in the ejecta morphology and thermodynamic properties. In addition to that, longer-term simulations would be valuable for more accurately analysing the evolution of neutrino-driven outflows and spiral wind ejecta.

Since disruptive black hole–neutron star mergers produce massive discs, including magnetic fields in future simulations would provide a more realistic picture of the remnant environment. It may also shed light on magnetic field amplification processes, which play a key role in launching jets and shaping electromagnetic counterparts.

\section{acknowledgments}
    RM would like to thank Beyhan Karakaş, Tim Dietrich, Francois Foucart, Moritz Reichert and Nikhil Sarin for valuable discussions, and David Radice for both valuable discussions and providing the solar abundance pattern data. We acknowledge the use of the IRIDIS High Performance Computing Facility, and associated support services at the University of Southampton. NA and IH also gratefully acknowledge support from the Science and Technology Facility Council (STFC) via grant numbers ST/R00045X/1 and ST/V000551/1.

\section{Data Availability}
We use a modified version of the parameter file described in~\cite{matur2024tab}. An example parameter file, along with a subset of the raw and processed data, is available on Zenodo~\citep{zenodospinning}.

\bibliographystyle{mnras}
\bibliography{sample}

\appendix
\section{Appendix}\label{appendix}

In this section, we examine the sensitivity of our $r$-process nucleosynthesis results to the number of tracer particles.

\begin{figure}
\centering
\includegraphics[width=0.5\textwidth]{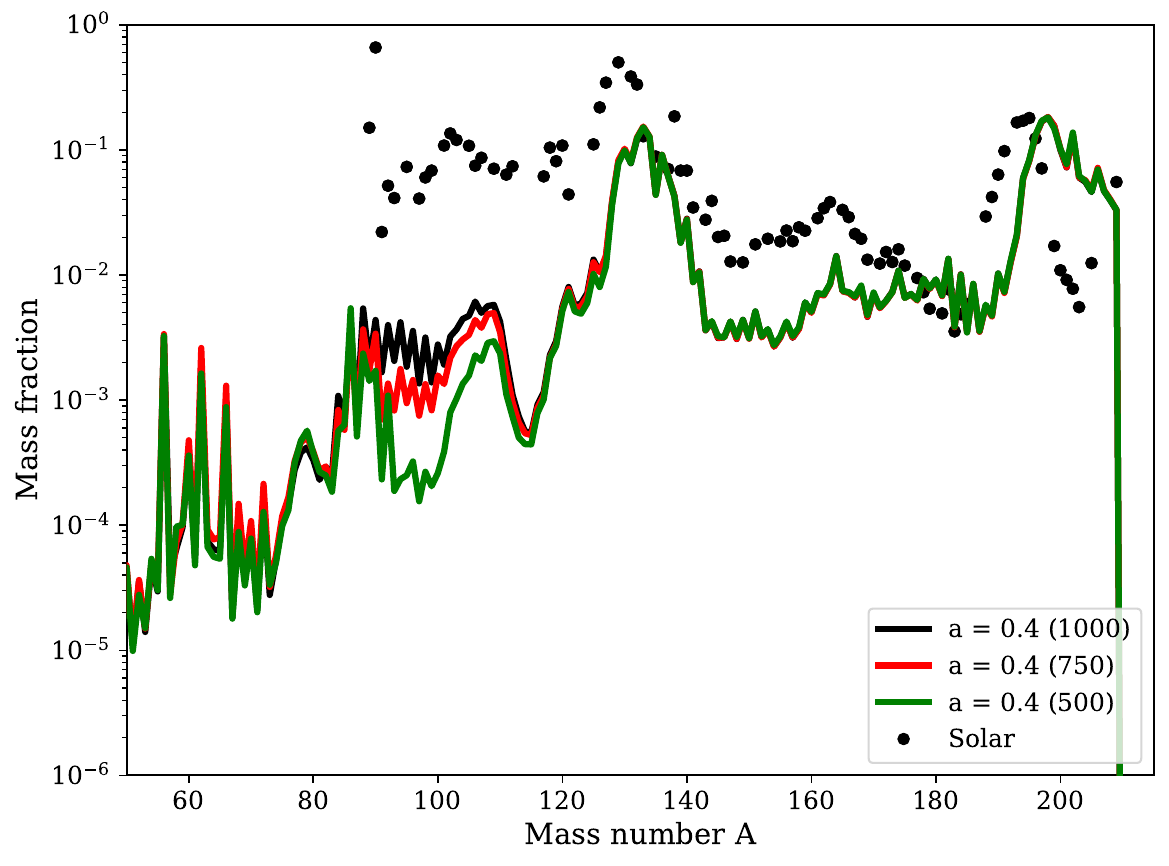}
\caption{Normalized $r$-process nucleosynthesis results for the \texttt{Q2.6a04} model using three different numbers of tracer particles: 500, 750, and 1000. As can be seen, the only noticeable difference appears in the range $90 < A < 110$, which corresponds to a very small mass fraction.} \label{fig:convergencerprocess}
\end{figure}

As mentioned in~\cref{sec:rprocess}, we perform this analysis using different numbers of tracer particles to examine whether the results change. Accordingly, we provide~\cref{fig:convergencerprocess} to illustrate the differences.

As can be seen from the figure, the only noticeable difference appears in the range $90 < A < 110$, which lies below the second peak and is not particularly important for \gls{BHNS} mergers, as $Y_{e}$ is significantly lower than in \gls{BNS} mergers. This figure demonstrates that the results remain nearly the same with increasing numbers of tracer particles.

\bsp
\label{lastpage}
\end{document}